\documentclass[twocolumn,preprintnumbers,amsmath,amssymb]{revtex4}

\usepackage{graphicx}
\usepackage{dcolumn}
\usepackage{bm}
\usepackage{epsfig}
\usepackage{amssymb}
\begin{document}

\preprint{}

\title{Comment on ``Search for Axions with the CDMS Experiment''}

\author{J.I. 
Collar$^{a}$ and M.G. Marino$^{b}$
}
\address{ 
$^{a}$Kavli Institute for Cosmological Physics and Enrico 
Fermi Institute, University of Chicago, Chicago, IL 60637\\
$^{b}$Center for Experimental Nuclear Physics and
Astrophysics and Department of Physics, University of Washington,
Seattle, WA 98195\\
}


\maketitle

The CDMS collaboration has recently reported new limits on axion-like 
particles \cite{cdms}, constraining the origin of the DAMA 
annual modulation effect. It is claimed in \cite{cdms} that 
 limits from both CDMS and CoGeNT \cite{cogent} 
``completely exclude the DAMA allowed region'' able to explain 
the modulation. This is in contrast 
with \cite{cogent}, where 
claims of DAMA exclusion are avoided, emphasizing instead 
a competitive DAMA sensitivity to pseudoscalars. 
We clarify here this possible source of confusion.

An interpretation of the DAMA modulation based on 
the axio-electric effect is prompted by the observation of
a peak in their spectrum at 3.2 keV, 
where the modulation is also maximal. A fraction of these events can arise from 
mundane $^{40}$K decays. $^{nat}$K is present in DAMA crystals at $<20$ 
ppb, the upper limit being able to explain the 
total of the $\sim0.75$ counts/kg-day 
under the peak. Several sources of uncertainty lead to a broad 
allowed region in axio-electric coupling {\frenchspacing vs.} pseudoscalar mass \cite{damapseudo}. 

\begin{figure}
\includegraphics[width=7.7cm]{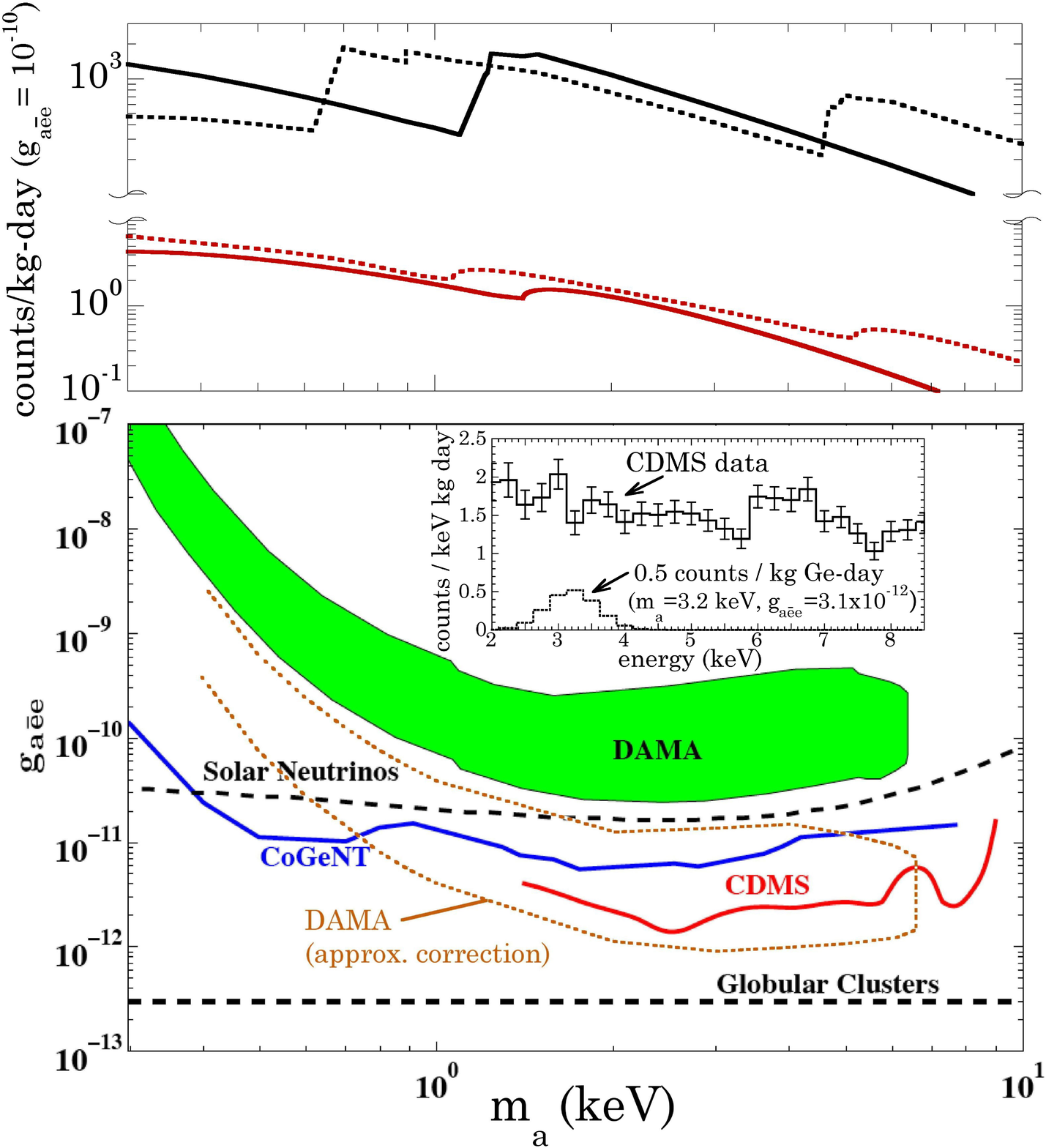}
\caption{\label{fig:epsart} Top: Axio-electric rates in germanium (solid) and NaI (dotted). 
Upper lines follow  
\protect\cite{pospelov}, lower 
follow \protect\cite{damapseudo}. Modern photoelectric cross-sections 
are used in the first case, a hydrogenic wavefunction 
approximation in the second. Bottom: Axio-electric exclusion 
boundaries from CoGeNT and CDMS, also displaying the DAMA allowed 
region (see text).
}
\end{figure}

Pospelov {\it {\frenchspacing et al.}} \cite{pospelov} pointed out 
two intimately related flaws in the definition 
of this allowed region. First, the leading term in the Hamiltonian for 
the interaction would have been left out in DAMA's derivation of the 
axio-electric rate \cite{damapseudo}. 
{\frenchspacing Fig. 1 (top)} shows the large differences in rate 
following the formula in \cite{pospelov} and  
calculations based on \cite{damapseudo}. 
The ratio 
of the NaI rates is used to estimate 
what would have been the correct location of the DAMA allowed region ({\frenchspacing 
Fig. 1, bottom}). This is the origin for the 
moderate statements in \cite{cogent}: while CoGeNT would impose a lower 
bound $m_{a}>$0.8 keV, neither CoGeNT nor CDMS would presently 
exclude all of the  phase space allowed
for a pseudoscalar interpretation of the DAMA effect.

The second objection in \cite{pospelov} is however more fundamental. General 
principles are cited that point 
to a typical $1/v$ dependence for the inelastic scattering 
cross-section of non-relativistic particles, as is the case here for 
the correct Hamiltonian. Flux being 
proportional
to $v$, the interaction rate should be (to first order) velocity-independent, 
negating the possibility of a modulation large enough to 
explain DAMA's observations.  In this situation it clearly makes no sense to speak 
of an allowed region. We instead expect 
the competitive DAMA sensitivity mentioned in \cite{cogent} 
(i.e.,  
an exclusion boundary comparable to that from CDMS and CoGeNT). To illustrate the 
point, the inset in {\frenchspacing Fig. 1}
shows the signal from a pseudoscalar responsible for 50\% of the intensity 
under the 3.2 keV DAMA peak. This is clearly at the limit of 
detectability for CDMS.

We conclude by noting that it may be possible to 
recover a source for the modulation and with it the concept of an 
allowed or favored DAMA region.
For instance, if a solar-bound population of dark 
pseudoscalars is invoked, yearly changes by few percent in 
interaction rate can arise from velocity-independent 
considerations: the Earth's orbital ellipticity modulates the phase space of their 
viable orbits \cite{solarbound}, with extrema in early July and January 
(either sign is possible depending on which orbital 
invariants are favored by the population). In other words, the 
modulation would
enter via the number density of the particles, not their velocities. 
A reasonable (i.e., beyond reach of existing bounds) 
local density for this population would 
bypass the remaining astrophysical constraints on 
their coupling (globular clusters, {\frenchspacing Fig. 1}). 
This possibility illustrates the rich phenomenology available for a 
dark matter interpretation of the DAMA effect.

We would like to thank Maxim Pospelov and Georg Raffelt for many 
useful exchanges. Further clarification and discussion on these 
issues can be found 
in \cite{georg}.

\end{document}